\documentclass[sn-mathphys]{sn-jnl}

\jyear{2022}%
\theoremstyle{thmstyleone}%
\theoremstyle{thmstyletwo}%
\theoremstyle{thmstylethree}%
 \usepackage{aas_macros}

\def\lesssim{\mathrel{\hbox{\rlap{\hbox{\lower4pt\hbox{$\sim$}}}\hbox{$<$}}}}
\def\gtrsim{\mathrel{\hbox{\rlap{\hbox{\lower4pt\hbox{$\sim$}}}\hbox{$>$}}}}

\def\micron{\hbox{$\mu$m}}

\definecolor{ao}{rgb}{0.0, 0.5, 0.0}

\newcommand\reduline{\bgroup\markoverwith{\textcolor{red}{\rule[-0.5ex]{2pt}{0.4pt}}}\ULon}

\newcommand{\Msun}{\hbox{M$_\odot$}}
\newcommand{\msol}{\hbox{M$_\odot$}}

\newcommand{\OVI}{[\hbox{{\rm O}\kern 0.1em{\sc vi}}]}

\newcommand{\NV}{\hbox{{\rm N}\kern 0.1em{\sc v}}}
\newcommand{\SiIV}{\hbox{{\rm Si}\kern 0.1em{\sc iv}}}
\newcommand{\OIV}{[\hbox{{\rm O}\kern 0.1em{\sc iv}}]}
\newcommand{\NIV}{[\hbox{{\rm N}\kern 0.1em{\sc iv}}]}
\newcommand{\CIV}{\hbox{{\rm C}\kern 0.1em{\sc iv}}}
\newcommand{\HeII}{\hbox{{\rm He}\kern 0.1em{\sc ii}\kern 0.1em{$\lambda1640$} }}
\newcommand{\OIII}{[\hbox{{\rm O}\kern 0.1em{\sc iii}}]{$\lambda5007$}}
\newcommand{\OIIId}{[\hbox{{\rm O}\kern 0.1em{\sc iii}}]{$\lambda4959\lambda5007$}}

\newcommand{\NIII}{[\hbox{{\rm N}\kern 0.1em{\sc iii}}]}
\newcommand{\AlIII}{\hbox{{\rm Al}\kern 0.1em{\sc iii}}}
\newcommand{\SiIII}{\hbox{{\rm Si}\kern 0.1em{\sc iii}}}
\newcommand{\CIII}{\hbox{{\rm C}\kern 0.1em{\sc iii}]}}
\newcommand{\NeIV}{[\hbox{{\rm Ne}\kern 0.1em{\sc iv}}]}
\newcommand{\MgII}{\hbox{{\rm Mg}\kern 0.1em{\sc ii}}}

\newcommand{\CII}{[\hbox{{\rm C}\kern 0.1em{\sc ii}]}}

\newcommand{\He}{\hbox{{\rm He}\kern 0.1em{\sc ii}\kern 0.1em{$\lambda1640\lambda4686$}}}

\newcommand{\Halpha}{H$\alpha$}
\newcommand{\Hbeta}{H$\beta$}
\newcommand{\Hgamma}{H$\gamma$}
\newcommand{\SII}{[\hbox{{\rm S}\kern 0.1em{\sc ii}}]$\lambda6717\lambda6731$}
\newcommand{\NII}{[\hbox{{\rm N}\kern 0.1em{\sc ii}}]}
\newcommand{\OII}{[\hbox{{\rm O}\kern 0.1em{\sc ii}}]}

\newcommand{\MgI}{\hbox{{\rm Mg}\kern 0.1em{\sc i}}}
\newcommand{\FeII}{\hbox{{\rm Fe}\kern 0.1em{\sc ii}}}

\newcommand{\OI}{\hbox{{\rm O}\kern 0.1em{\sc i}}}
\newcommand{\NeII}{[\hbox{{\rm Ne}\kern 0.1em{\sc ii}}] }
\newcommand{\NaI}{[\hbox{{\rm Na}\kern 0.1em{\sc i}}] }
\newcommand{\NeIII}{[\hbox{{\rm Ne}\kern 0.1em{\sc iii}}] }

\begin{document}

\title[$z\sim3-4$ Quiescent galaxies with JWST]{A population of faint, old, and massive quiescent galaxies at $3<z<4$ revealed by JWST NIRSpec Spectroscopy}

\author*[1]{\fnm{Themiya} \sur{Nanayakkara}}\email{wnanayakkara@swin.edu.au}
\author[1]{\fnm{Karl} \sur{Glazebrook}}
\author[1]{\fnm{Colin} \sur{Jacobs}}
\author[1]{\fnm{Lalitwadee} \sur{Kawinwanichakij}}
\author[2]{\fnm{Corentin} \sur{Schreiber}}
\author[3]{\fnm{Gabriel} \sur{Brammer}}
\author[1]{\fnm{James} \sur{Esdaile}}
\author[1]{\fnm{Glenn G.} \sur{Kacprzak}}
\author[1]{\fnm{Ivo} \sur{Labbe}}
\author[3,10,11]{\fnm{Claudia} \sur{Lagos}}
\author[4]{\fnm{Danilo} \sur{Marchesini}}
\author[5]{\fnm{Z.~Cemile} \sur{Marsan}}
\author[3,6]{\fnm{Pascal A.} \sur{Oesch}}
\author[7]{\fnm{Casey} \sur{Papovich}}
\author[8]{\fnm{Rhea-Silvia} \sur{Remus}}
\author[9,10,12]{Kim-Vy H. Tran}


\affil*[1]{\orgdiv{Centre for Astrophysics and Supercomputing}, \orgname{Swinburne
  University of Technology}, \orgaddress{\street{P.O. Box 218}, \city{Hawthorn}, \postcode{3122}, \state{VIC}, \country{Australia}}}
  
\affil[2]{\orgname{IBEX Innovations},  \orgaddress{\street{Sedgefield}, \city{Stockton-on-Tees}, \postcode{TS21 3FF}, \country{United Kingdom}}}

\affil[3]{\orgdiv{Cosmic DAWN Center, Niels Bohr Institute}, \orgname{University of Copenhagen},  \orgaddress{\street{Jagtvej 128}, \city{Copenhagen N}, \postcode{DK-2200}, \country{Denmark}}}

\affil[4]{\orgdiv{Physics and Astronomy Department}, \orgname{Tufts University}, 
\orgaddress{574 Boston Avenue}, \city{Medford}, \state{MA}, \postcode{02155}, \country{USA}}

\affil[5]{\orgdiv{Department of Physics and Astronomy}, \orgname{York University}, 
\orgaddress{4700 Keele Street}, \city{Toronto}, \state{ON}, \postcode{M3J 1P3}, \country{Canada}}

\affil[6]{\orgdiv{Department of Astronomy}, \orgname{University of Geneva},  \orgaddress{\street{Chemin Pegasi 51}, \city{Versoix}, \postcode{CH-1290}, \country{Switzerland}}}

\affil[7]{\orgdiv{Department of Physics and Astronomy, and George P. and Cynthia Woods Mitchell Institute for Fundamental Physics and Astronomy}, \orgname{Texas A\&M University},  \orgaddress{\city{College Station}, \state{TX} \postcode{77843-4242}, \country{USA}} }

\affil[8]{University Observatory Munich, Faculty of Physics, Ludwig-Maximilians-University, Scheinerstrasse 1, 81679, Munich, Germany}

\affil[9]{School of Physics, University of New South Wales, Kensington, Australia}
\affil[10]{ARC Centre for Excellence in All-Sky Astrophysics in 3D}

\affil[11]{\orgdiv{International Centre for Radio Astronomy Research}, \orgname{University of Western Australia}, \orgaddress{\street{7 Fairway}, \city{Crawley}, \postcode{6009}, \state{WA}, \country{Australia}}}

\affil[12]{\orgdiv{Center for Astrophysics}, \orgname{Harvard \& Smithsonian}, \orgaddress{\city{Cambridge}, \state{MA}}}

\maketitle

\section{Abstract}\label{abstract}

Here we present a sample of 12 massive quiescent galaxy candidates at $z\sim3-4$ observed with the James Webb Space Telescope (JWST) Near Infrared Spectrograph (NIRSpec). These galaxies were pre-selected from the \emph{Hubble} Space Telescope imaging and 10 of our sources were unable to be spectroscopically confirmed by ground based spectroscopy. By combining spectroscopic data from NIRSpec with multi-wavelength imaging data from the JWST Near Infrared Camera (NIRCam), we analyse their stellar populations and their formation histories. We find that all of our galaxies classify as quiescent based on the reconstruction of their star formation histories but show a variety of quenching timescales and ages. 
All our galaxies are massive ($\sim 0.1-1.2\times10^{11}$ \Msun), with masses comparable to massive galaxies in the local Universe. We find that the oldest galaxy in our sample formed $\sim 1.0\times10^{11}$ \Msun\ of mass within the first few hundred million years of the Universe and has been quenched for more than a billion years by the time of observation at $z\sim 3.2$ ($\sim2$ billion years after the Big Bang). Our results point to very early formation of massive galaxies requiring a high conversion rate of baryons to stars in the early Universe.

\section{Introduction}\label{intro}

The confirmation of the presence of massive ($\gtrsim10^{10}$ \msol) quiescent galaxies at epochs only 1–2 Gyr after the Big Bang \citep{Glazebrook2017,Marsan2017,Schreiber2018b,Tanaka2019,Carnall2020a,Forrest2020a,Forrest2020b,Valentino2020} has challenged models of cosmology and galaxy formation \citep{Merlin2019a}. 
These sources likely had extreme star-formation levels to build up the $\sim 10^{11}$ stellar masses within the first billion years of the Universe. 
They further require mechanisms that led to cessation of star-formation within very short timescales, while most of other galaxies were actively forming stars contributing to the growth of the cosmic star-formation rate density \citep{Madau2014}. 
Therefore, producing sufficient numbers of these requires abundant numbers of the host dark matter halos to have been assembled and sufficient time for star formation to proceed extremely quickly and then cease just as rapidly. 
Currently, the mechanisms that led to the rapid mass buildup and abrupt quenching of star-formation is an outstanding question.

The extremely rapid formation makes massive quiescent galaxies ideal laboratories to probe the most extreme galaxy formation scenarios in the early Universe and mechanisms that shut down star- formation. 
Ground-based spectroscopy has suggested ages of 200--300 Myr \citep{Schreiber2018b} at redshifts $3<z<4$.
The true number and ages of these objects have however been highly uncertain as ground-based spectra has been limited to the brightest of them \citep[e.g.][]{Schreiber2018b,Carnall2020a}, at wavelengths $\sim 2~\micron $ , which introduces a significant potential bias towards younger objects \citep{Forrest2020b}. 
Additionally, Balmer or D4000 breaks often fall between atmospheric transmission windows at $<2~\micron$, limiting constraints to age/formation timescales of these galaxies \citep[e.g.][]{Nanayakkara2021}.

Short lived extreme star-formation episodes required to produce these galaxies may have implications for cosmological and chemical evolution models. 
Massive stars prominent in low metallicity environments \citep{Narayanan2012}, such as in the early Universe, would drive galaxy chemical enrichment through core-collapse supernovae \citep{Nomoto2006a} leading to an enhancement of $\alpha-$ elements in the inter stellar medium and stars \citep[e.g.][]{Kriek2016}.
$\alpha-$enhanced stars are likely to be more massive (leading to a top-heavy IMF), have less Fe blanketing, and weaker stellar winds \citep{Steidel2016}.
These effects lead to an enhancement of ionizing photons per unit star-formation rate. 
Additionally, core collapse supernovae leads to stronger feedback in galaxies, creating low column density channels for ionizing photons to escape. Therefore, these sources could have played a dominant role in the reionization of the Universe \citep{Naidu2019}. Stronger feedback would also contribute to shut down star-formation \citep{Kimmig2023a}. These two effects could lead to changes in the reionization timescales posing newer challenges in cosmology and supernovae chemical enrichment. Thus, it is imperative to explore new mechanisms that should be considered to efficiently build up mass and cease star-formation in this very short time window.

Constraining the abundance of massive quiescent galaxies at $z>3$ and the nature of their stellar population is important to provide constraints for  galaxy evolution models.
Many simulations currently investigate possible impacts of modified Active Galactic Nucleus (AGN) feedback. \citep[e.g., Illustris-TNG, Magneticum, SHARKS \& Eagle][]{Kurinchi-Vendhan2023a,Lagos2023a,Remus2023a}  and different depletion timescales and star formation models \citep{Valentini2023a}  on the star formation and quenched properties of high redshift galaxies. 
With current observations, we have a very limited understanding on the formation pathways of these galaxies, thus, observations of $z>3$ massive quiescent galaxies are crucial to refining our understanding of the complex balance of physical processes that play a key role in the early epochs of galaxy formation.
While it is crucial to determine the nature of AGN in the $z>3$ massive quiescent galaxies, at $z>3$ rest-optical AGN diagnostics \citep[e.g.][]{Baldwin1981} cannot be obtained due to atmospheric cutoff at $\gtrsim 2.5~\micron $ from ground based spectroscopy.

The launch of the James Webb Space Telescope (JWST) enables dramatically more sensitive and constraining spectroscopic observations due to the very low sky background, sharp image quality, and access to wavelengths beyond 2 \micron. 
The first observations have already revealed new photometric candidates for massive quiescent galaxies at $z>3$ \citep[e.g.][]{Carnall2023a,Long2023b} and even their possible progenitors at $z\sim9$ \citep{Labbe2022a}.

Here we report JWST NIRSpec \citep{Jakobsen2022a}  (0.6--5.3 \micron) observations of 12 quiescent galaxy candidates at $z\approx3-4$, out of which 10 were beyond the limit of previous ground-based spectroscopy. 
Our objects were selected from the sample of Schreiber+2018 \citep[henceforth S18][]{Schreiber2018} who compiled a deep complete spectroscopic analysis of massive quiescent galaxy candidates at $z>3$. These were obtained using rest-frame $U-V$ vs $V-J$ color selection techniques with pre-JWST data \citep{Williams2009}.  
The prior ground-based spectra covered wavelengths 1.5--2.3 \micron\ and were taken with the MOSFIRE spectrograph on the 10~m Keck telescope  \citep{McLean2008} with up to 14 hours on-target exposures.
Out of the 24 massive galaxy candidates observed by S18, spectroscopic redshifts were obtained for only 12 of the galaxies. 
The majority of the confirmed objects were found to have redshifts $z>3$, where only two were low-redshift interlopers.  
The spectroscopically confirmed $z > 3$ galaxies were all typically brighter than $K=23$, as the observations are limited by the bright terrestrial $K$-band background. 
This potentially introduces a significant age bias. 
At these redshifts the observed $K$-band probes the rest frame $B$-band. 
At fixed mass, younger objects will have smaller $B$ band mass to light (M/L) ratios, hence be brighter; older objects, conversely, will be fainter, being dominated by older stellar populations, with higher M/L ratios. These older objects, having formed earlier, would be more constraining on theoretical models \citep{Forrest2020b}.


\section{Results}\label{results}

On 1$^{\hbox{st}}$ of Aug, 2022 using JWST NIRSpec we obtained spectra of eight massive quiescent galaxy candidates from S18 that had eluded ground-based spectroscopic confirmation over the UKIDSS Ultra-Deep Survey field \citep[UDS,][]{Skelton2014,Straatman2016}. 
Additionally, due to close clustering we were able to include ZF-UDS-8197, a galaxy that was confirmed by S18 in one of our configurable microshutter array (MSA) configurations \citep{Rawle2022a}. 
On the 11$^{\hbox{st}}$ the of April, 2023 we further obtained spectroscopy for 2 massive quiescent galaxy candidates from S18 over the All-Wavelength Extended Growth Strip International Survey field \citep[EGS,][]{Skelton2014}. Similar to UDS, due to close clustering we were able to include 3D-EGS-18996 which was previously confirmed by S18.

\begin{figure}[h]%
\includegraphics[scale=0.2]{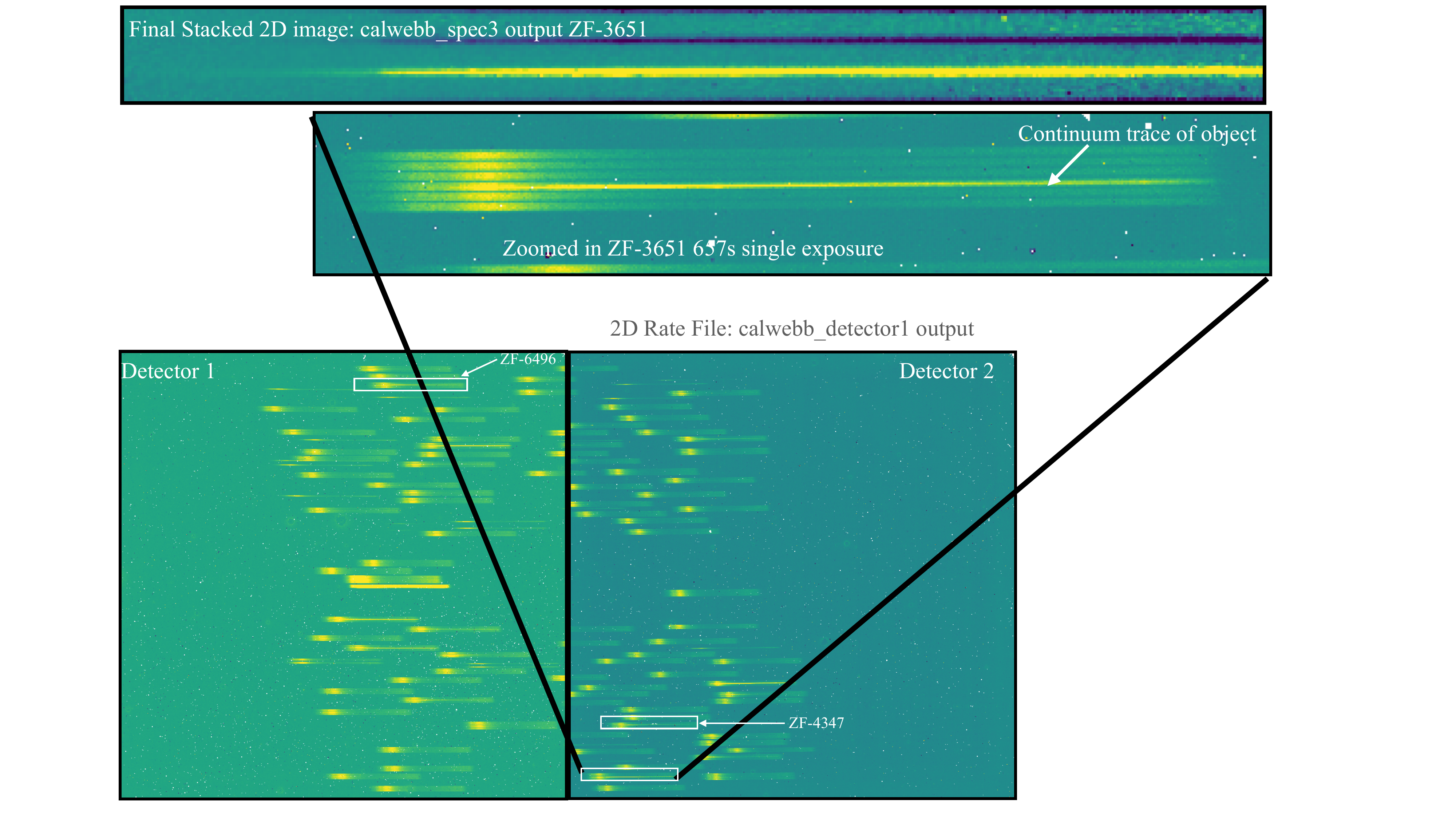}
\caption{An example of raw and reduced data obtained by our program. The lower panels show the first exposure of our ZFOURGE UDS mask {\tt obs100}. The data has been processed through the {\tt calwebb\_detector1} pipeline which has combined the raw frames of the exposure (ramp fitted). This was a 657~s exposure (similar to all exposures carried out by our program) and three of our quiescent candidates: ZF-UDS-4347, ZF-UDS-3651, and ZF-UDS-6496 were covered by this pointing. We have highlighted the positions of these three galaxies in the detector using white boxes.  The continuum traces from our $z>3$ galaxies can be clearly seen in this exposure. In the middle panel we have zoomed into the region in the detector which covers ZF-UDS-3651.  The continuum of the object and the bar shadows from the MSA can be seen here. The upper panel shows the final reduced 2D image of ZF-UDS-3651. Three exposures of 657~s in three dither positions have been combined to produced the positive trace of this object which is optimally extracted and shown in Figure \ref{fig:spectra_noinset}. 
\label{fig:spectra_2D}}
\end{figure}

Our observations used the MSA to form programmable slits and the low-resolution ($50<R<500$) prism disperser with the CLEAR filter covering wavelengths  0.6--5.3~\micron. 
Here we present results from four MSA configurations ({\tt obs6}, {\tt obs100}, {\tt obs200}, {\tt obs300}; these configurations are selected because most of the sources overlap with JWST PRIMER-UDS imaging (GO-1837, PI Dunlop) and JWST CEERS imaging \citep[DD-ERS-1345, PI Finkelstein][]{Finkelstein2023a}). 
We used 5 slitlet shutters with 3 dither positions. Each dither position was observed for 657~s. 
In Figure \ref{fig:spectra_2D} we show the 2D raw and reduced frames of one of our MSA configurations, {\tt obs100}. 
Three primary targets fall in this mask and the continua of these objects are clearly visible in the ramp-fitted raw frames.

The absolute flux calibration accuracy of data reduced with post launch calibration files is expected to be at $\lesssim5\%$ level (STScI, private communication). 
However, the STScI pipeline currently does not account for slit loss functions accurately for NIRSpec MOS extended sources. Therefore we used multi-band NIRCam imaging from the PRIMER survey to apply an empirical scaling to the data as follows.
For each object, we selected all photometric bands within $1.0~\mu m-5.0~\mu m$ detected with a signal-to-noise ratio (S/N) of $>20$ and computed the offset between the flux measured from the NIRSpec spectrum and broadband/medium-band photometry. We then fit a 2nd order polynomial to the offset as a function of wavelength to obtain a spectrophotometric scaling factor for each galaxy.
The order of the polynomial was chosen carefully via visual inspection for each object, to verify no artificial structure or colour term would be introduced to the spectra as a result of this process. 
We multiplied the observed spectra by this calibration function to obtain NIRSpec spectra which are consistent with the PRIMER photometric data.
2 galaxies, 3D-UDS-39102 and 3D-UDS-41232 fall outside the PRIMER imaging area, thus, photometric data with a S/N $>10$ from the 3D-HST survey \citep{Skelton2014} is used instead. 
More details on the scaling process and associated tests performed to verify the calibration accuracy of the STScI pipeline are provided in the Methods Section \ref{sec:spec_cal}.

\begin{figure}[h]%
\includegraphics[trim={10 0 0 0},clip, scale=0.50]{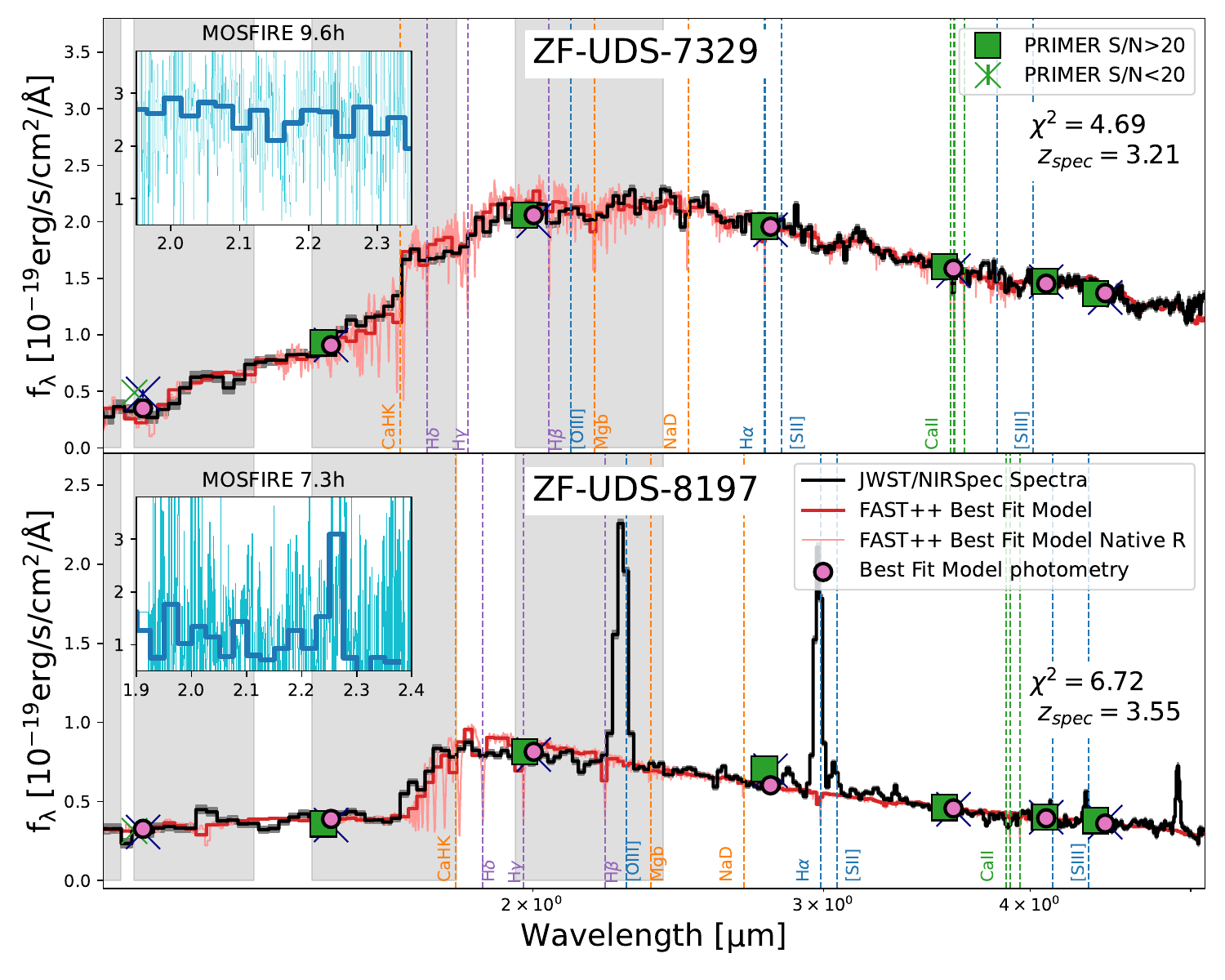}
\caption{The {\bf top} and {\bf bottom} panels present JWST/NIRSpec spectra of two quiescent galaxy candidates, ZF-UDS-7329 and ZF-UDS-8197 respectively, observed over the ZFOURGE-UDS field. 
ZF-UDS-7329 is one of the oldest known quiescent galaxies in the $z>3$ Universe \citep{Glazebrook2023a}. ZF-UDS-8197 is a quiescent galaxy in our sample that exhibits strong \OIII\ and \Halpha\ EWs, possibly driven by an AGN. The spectra were optimally extracted \citep{Horne1986} and flux-calibrated using PRIMER photometry. Grey bands highlight the KECK/MOSFIRE $Y$, $J$, $H$, $K$ bands (from left to right). 
To improve clarity, spectra are trimmed at $<1.1~\mu m$. 
For each galaxy, we also display the PRIMER total photometry in two bins based on the S/N. 
The best-fit {\tt FAST++} template and the best-fit model photometry for the observed filters used in {\tt FAST++} are shown in the panels. 
Commonly observed rest-frame emission and absorption features are marked in the spectra. Insets in each panel display the ground-based $K$-band MOSFIRE spectra presented in S18. 
The thin cyan lines represent the MOSFIRE spectra at its native resolution of $R\sim3000$, while the thick blue lines correspond to the MOSFIRE spectra at a resolution similar to that of the JWST/NIRSpec Prism observations at $\lambda_{obs}\sim 2~\mu m$ ($R\sim100$).
The best-fit redshift and the reduced $\chi^2$ of the {\tt FAST++} fits are labelled in each panel.
\label{fig:spectra_inset}}
\end{figure}

\begin{figure}[h]%
\includegraphics[trim={10 0 0 0},clip, scale=0.35]{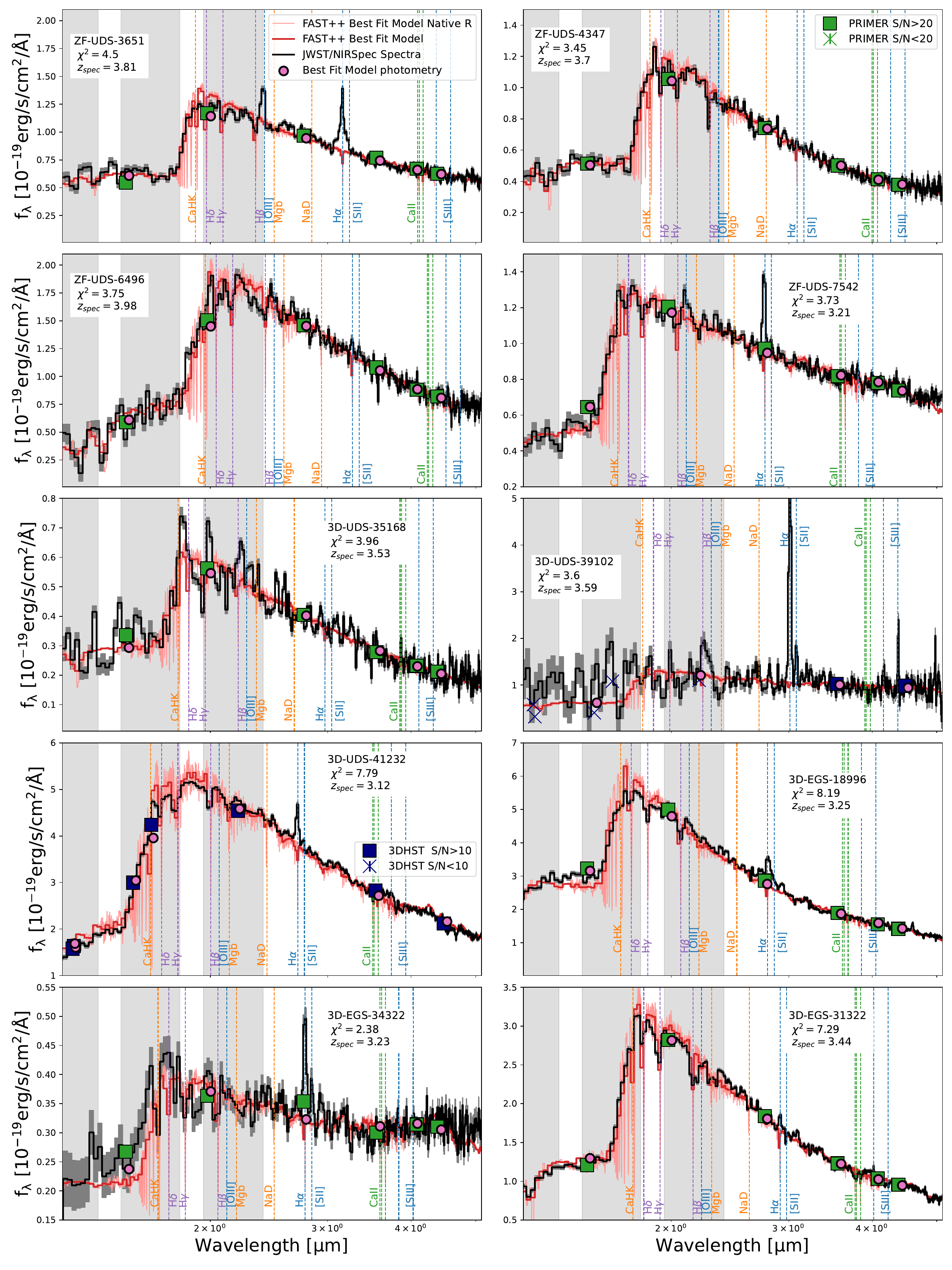}
\caption{
JWST/NIRSpec spectra of the other quiescent galaxy candidates presented in this work. The format of this figure mirrors that of Figure \ref{fig:spectra_inset}, with the exception of the MOSFIRE insets.
\label{fig:spectra_noinset}}
\end{figure}

In Figures \ref{fig:spectra_inset} and \ref{fig:spectra_noinset} we show the calibrated NIRSpec spectra for our galaxies observed over the UDS and EGS fields. 
11 galaxies show a strong Balmer break confirming their post-starburst nature. 
The Balmer and D4000 break of a galaxy is indicative of the nature of stellar populations dominating the continuum \citep[e.g.][]{Poggianti1997}. 
Once the young O stars move off the main sequence, the galaxy continuum is dominated by late-B and main sequence A stars which gives rise to a Balmer break. The strength of this feature will reduced after a few hundred million years from the most recent star-formation episode due to A type stars moving away from the main sequence. 
Thus, at older ages, absorption from ionized metals from old late type stars' atmospheres will dominate the absorption features at $\sim4000$~\AA. 
In the early Universe, due to the limited lifetime for galaxy evolution, Balmer break galaxies will dominate over D4000 break galaxies. 
JWST NIRSpec/Prism observations are expected to uncover large populations of Balmer break galaxies in the early Universe \citep[e.g.][]{Binggeli2019a}.

One of our galaxies, ZF-UDS-8197, shows very prominent emission lines.  
Based on its $\rm{log_{10}}$(\OIII/\Hbeta) flux ratio of $1.4$, we conclude that an AGN is powering the emission lines of this galaxy \citep{Juneau2014a}.
There are further 7 galaxies which either show \OIII\ or \Hbeta\ flux  detections with a S/N$>3$ for which we can obtain limits to their $\mathrm{log_{10}}$(\OIII/\Hbeta) flux ratios. At their respective stellar masses they lie in the AGN or AGN and star-forming composite region based on the Mass Excitation (MEx) diagram \citep{Juneau2014a}. 
However, at $z>2$ the evolution of the ionization parameter \citep[e.g.][]{Steidel2014} adds further complexity to the interpretation of this diagram, specially for sources in the AGN and star-forming composite region.
Thus, we cannot rule out underlying star-formation contributing to the emission lines of these 7 sources. 
At NIRSpec/Prism resolutions, \Halpha\ emission is blended with the \NII\ doublet, thus we obtain lower limit of SFR$<13$~\msol/yr for the remaining four sources. 

With the exception of 3D-EGS-18996, the continua of all galaxies are detected at a median S/N of $>18$. 
While the resolution of NIRSpec/Prism observations is too low to constrain detailed element abundant patterns through absorption lines, some galaxies such as ZF-UDS-4347 show spectral signatures of hydrogen absorption such as \Hbeta\ and \Hgamma.
Dusty interlopes at $z\sim2$ can contaminate photometric selection of quiescent galaxies \citep{Antwi-Danso2022a}, however, with our NIRSpec/Prism observations we can confirm that none of our photometrically selected galaxies are $z\sim2$ dusty interlopers.

In Figure \ref{fig:spectra_inset} we also show the ground based Keck/MOSFIRE \citep{McLean2012} spectra obtained for two of the galaxies to compare with our new JWST observations.\textbf{} 
The native resolution of MOSFIRE is $R\sim3000$. S18 binned the spectra to $\sim10$ resolution elements to increase the continuum S/N. 
Given our NIRSpec/Prism observations have a much lower resolution of $R\lesssim100$ at $\lambda_{obs}\sim2~\mu m$ (where the Balmer break falls), we rebin the MOSFIRE spectra to 250 \AA\ bins to accurately compare with our NIRSpec observations.

ZF-UDS-8197 has strong \OIIId\ doublet detection that falls within the MOSFIRE-$K$ band. Thus, S18 obtained a confident redshift confirmation with an on-source exposure time of $\sim7$~h reaching an emission line S/N of $\sim12$. 
Even though MOSFIRE observations reached a continuum S/N of $\sim16$ with 70~\AA\ spectral binning, it was not possible to obtain signatures of absorption lines. 
Given the Balmer break largely falls within the atmospheric cutoff between $H$ and $K$ bands, it was not possible to confirm the post-starburst nature of this galaxy. 
However, with our new NIRSpec observations we are able to observe the Balmer break and other strong emission lines such as \Halpha\ and \SII.

ZF-UDS-7329 was observed for $\sim10$~h in $K$-band with MOSFIRE reaching a continuum S/N of $\sim17$. However, S18 was unable to obtain a spectroscopic redshift confirmation for this source due to the limited wavelength coverage.
Our NIRSpec observations show the clear spectral shape of this object and our {\tt slinefit} (https://github.com/cschreib/slinefit) redshift estimate puts it at $z=3.207^{+0.002}_{-0.001}$. 
The Balmer break of this galaxy is smoother compared to the other galaxies in our sample and the spectral features have transitioned to a D4000 break suggesting a much older underlying stellar population \citep{Bruzual1983}. 
However, degeneracies between dust and star formation history (SFH) can make the interpretation difficult. 
Thus, we use {\tt FAST++} \citep{Schreiber2018b} to perform full spectral fitting of our galaxies to constrain their SFHs.

We simultaneously fit NIRCam photometric data (data from the 3D-HST survey is used for 3D-UDS-39102 and 3D-UDS-41232) with NIRSpec spectra together in {\tt FAST++} keeping redshifts fixed at their new spectroscopic values.
We use BC03 \citep{Bruzual2003} stellar population models with a Chabrier \citep{Chabrier2003} IMF and follow a Calzetti \citep{Calzetti2000} dust prescription model.  
We allow the metallicity to vary between 20\% to 250\% $Z_\odot$. 
We mask out strong emission lines in the observed spectra and follow the same prescription as defined by S18 to parameterize the SFHs. 
To account for the low and non-linear spectral resolution and dispersion of the NIRSpec/Prism mode observations, we compute empirical line spread functions (LSF) for our galaxies based on the PRIMER NIRCam images. As outlined in \citep[][Section 2.3]{Glazebrook2023a}, for each galaxy we compute the 1D slit profile across each of the 7 PRIMER NIRCam bands and multiply with the NIRSpec/Prism dispersion function to obtain a measure of the LSF. 
This profile is input into {\tt FAST++} as the {\tt SPEC\_LSF\_FILE} option, which is used by {\tt FAST++} to convolve the BC03 models to match with the observed NIRSpec resolution. 
The errors of all parameters are computed as the 1-$\sigma$ distribution of the best-fit values of 1000 Monte-Carlo iterations. 
More details of the spectrophotometric fitting process is outlined in methods Section \ref{sec:sed_fitting} and in  \cite{Glazebrook2023a}. 
The {\tt FAST++} best-fit results are presented in Table \ref{tab:fastparams}.

With the exception of 3D-EGS-34322, all other galaxies show {\tt FAST++} SFRs$_{10}$ of $<1.5$~\msol/yr and the stellar masses range between $\sim0.1-1.2\times 10^{11}$~\msol. 
This can be compared with a characteristic stellar mass of $\sim0.5\times10^{11}$~\msol\ expected at $z\sim3$ \citep{Tomczak2014}.
A majority of our galaxies have relatively low levels of dust attenuation ($A_v\lesssim0.4$). 
3D-EGS-34322 and 3D-UDS-39102 show the highest amount of dust with $A_v=1.8$ suggesting a non-negligible amount of dust in some of these early massive systems. 
In Figures  \ref{fig:spectra_inset} and \ref{fig:spectra_noinset} we also show the best-fit (lowest-$\chi^2$) spectrum for our galaxies. All galaxies are well constrained by the parameters used in our fitting. 
We have extensively worked with the data reduction to constrain wiggles in the observed spectra and determine features that are not well reproduced by stellar populations models. This is further detailed in \cite{Glazebrook2023a} and a full analysis of the our GO-2565 sample will be presented in Nanayakkara et al. (in prep) once the wider COSMOS-WEB Cycle 1 GO program is concluded.

Before JWST/NIRSpec was available, the limited spectral coverage from ground based spectroscopy and uncertain redshifts purely based on photometric data made it challenging to constrain the formation timescales. 
With spectroscopy we are able to obtain a redshift measurement and to accurately constrain the shape of the SEDs for galaxies without a very sharp Balmer break, (i.e., the more evolved objects). 
Furthermore, we are able to quantify any emission line contamination to the photometry (which can produce larger rest-frame optical breaks than the stellar continuum) which could result in an overestimate of the stellar masses. 
Therefore, with joint spectrophotometric fitting with {\tt FAST++} we are able to tightly constrain galaxy formation and quenching timescales.

\begin{figure}[h]%
\includegraphics[scale=.6]{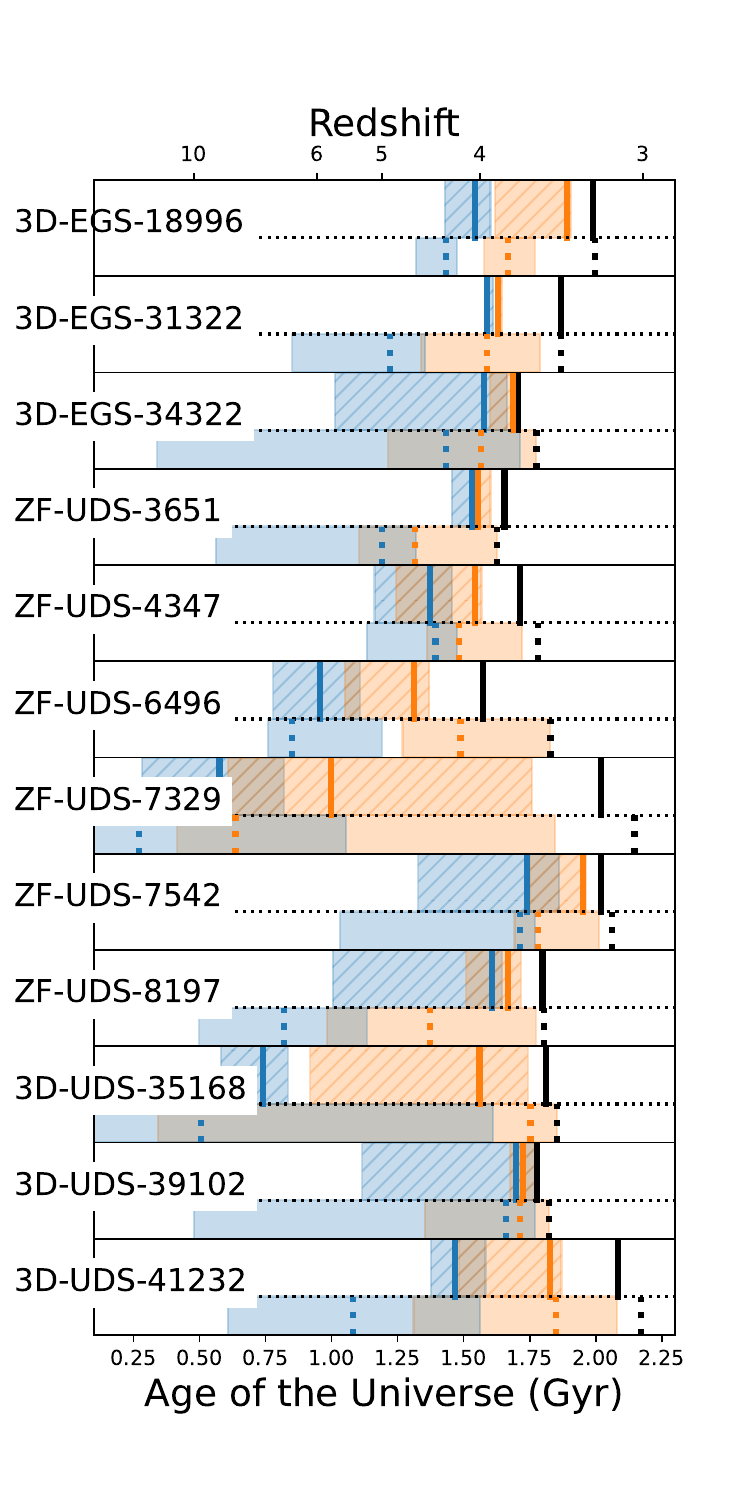}
\centering
\caption{The best-fit SFHs of our galaxies. The blue vertical lines indicate the time at which each galaxy formed 50\% of its stellar mass, with the associated $3\sigma$ error parameterized by 1000 MCMC iterations is depicted by the blue shaded region. We show the quenching time, as defined by S18 (the time at which the galaxy's Star Formation Rate (SFR) falls below $10\%$ of its primary SFR episode, as detailed in Section 4.1 of S18), by orange vertical lines. The associated error is shaded in orange. The black vertical line represents the age of the Universe when the galaxy is being observed. For each galaxy, the top panels depict the improved constraints acquired through our JWST/NIRSpec observations. The lower panels (below the dotted lines) present constraints reported in S18. For clarity, we use vertical dashed lines to represent the best-fit S18 values.
\label{fig:sfhs}}
\end{figure}

Figure \ref{fig:sfhs} shows the reconstructed SFHs based on {\tt FAST++} best fitting results. 
To be consistent with \cite{Glazebrook2023a}, we have trimmed the spectra to rest-frame $\sim0.7~\mu m$ in {\tt FAST++}  fitting for this purpose.
10 of our sources have updated JWST/NIRCam based photometry. This combined with the rest-frame optical spectral coverage from NIRSpec provide more accurate constraints to the reconstruction of the SFH for our sources compared to what was possible in S18.    
At the redshift of observation 3D-UDS-39102 is the youngest galaxy in our sample reaching $50\%$ of the stellar mass only in the last $\sim50$ million years. 
At the opposite end ZF-UDS-7329 reached its $50\%$ of the stellar mass $\sim1.5$ billion years before it is observed redshift of $z\sim3.2$. 
While all galaxies classify as quenched based on the S18 definition, it is clear that there is a large variety in quenching timescales. 
3D-EGS-34322 has only reached the $0.1\times \langle SFR \rangle_{main}$ quenching definition very recently in the last 20 million years, while ZF-UDS-7329 has been quenched for $\gtrsim1$Gyr. 
Once rest $U-V$ and $V-J$ colors are recomputed for our sample at their spectroscopic redshifts with strong emission line contributions removed \citep{Antwi-Danso2022a}, we find that all galaxies lie inside or relatively near the border of the quiescent region. 
ZF-8197 was previously observed in S18 and was ruled out as a star-forming galaxy based on rest $U-V$ and $V-J$ colors, however with newer constraints from JWST, we confirm its balmer break and find that it falls at the boundary between star-forming and quiescent.

Are such high numbers of massive quiescent galaxies plausible at $z\sim3-4$? 
As shown by several studies (\citep[e.g][]{Valentino2023a,Long2023a,Lustig2023a}) cosmological simulations found it challenging to reproduce observed number densities of massive quiescent galaxies at $z>3$. 
With our current observing program we have targeted sources selected in the pre-JWST era which eluded ground based spectroscopic confirmations and confirmed seven new quiescent galaxies.

S18 reported 20 $z\sim3-4$ massive quiescent galaxies in the full ZFOURGE field. 
This was based on an accuracy of 80\% (5/24 galaxies were reported as non-quiescent) for non-spectroscopically confirmed candidates. 
We observe 10 new galaxies and spectroscopically confirm seven to be quiescent. 3D-EGS-34322, ZF-UDS-7542, and 3D-UDS-39102  only reached the quiescent definition in the $\lesssim 100$ Myrs, so to be conservative, we don't consider them to be quiescent in our calculations. 
We find that ZF-UDS-8197 which was ruled out by S18 to be quiescent. 
Additionally, 3D-UDS-39102 while quiescent, has a stellar mass $<3\times10^{10}$~\msol. So we remove this also as an outlier since it doesn't satisfy the massive criteria imposed by S18. 
Thus our updated number density of massive quiescent sources is $\mathrm{\sim 1.1~(\pm 0.3) \times 10^{-5}~Mpc^{-3}}$.

On average, cosmological simulations have been able to reproduce observed number densities of quiescent galaxies up to $z\sim3$ \citep{Valentino2020,Long2023a,Lustig2023a}, and depending on environment even up to $z=4$ \citep{Remus2022a}. 
For example our values can be compared with $\mathrm{0.6-3.6\times10^{-5}~Mpc^{-3}}$ obtained respectively at $z=3.7$ and $z=3$ with Illustris-TNG 100 \citep{Valentino2020}. 
SHARK simulations also find a similar number density evolution for quiescent galaxies at $z\sim3-4$ \citep{Long2023a}.
Additionally, at $z\sim4$, the local environment of galaxies has also shown to play a role in determining the quiescence of galaxies \citep{Remus2022a}.

In order to reconcile observed number densities of massive quiescent galaxies at $z\sim3-4$, simulations have explored  how supermassive black hole feedback regulate galaxy growth and star-formation quenching in galaxies in the first $1-2$ billion years of the Universe and effects of revising gas/dust depletion timescales and star burst properties of the early Universe \citep{Hartley2023a,Kimmig2023a,Lagos2023a,Remus2023a,Valentini2023a}.
Observations such as those presented in this analysis will be critical to refining our understanding of the complex balance of physical processes that play the key role in the early epochs of galaxy formation.

Recent results from TNG300 show that with modified AGN feedback models that kick in at $z<6$, currently observed number densities of massive quiescent galaxies can be reproduced \cite{Hartley2023a}. Similarly, SHARK v2.0 semi-analytical model is also able to reproduce number densities largely in agreement with observed values at $z<5$ \citep{Lagos2023a}.   
However, details such as ages are still in tension and under investigation from several simulation approaches. 
For example, TNG300 forms the first quiescent galaxies at $z<4.2$ and as shown by \citep{Glazebrook2023a}, the early formation of ZF-UDS-7329 is still a challenge for TNG300.
In contrast to TNG300, the higher simulation box size and the greater time resolution of the Magneticum Pathfinder Simulations is attributed to Magneticum simulations being able to produce massive quenched galaxies at a higher redshift \cite[from $z\sim7$ onward ][]{Kimmig2023a,Remus2023a}

At the end of our program we will provide tighter constraints to the abundance and formation epochs of massive quiescent galaxies in the early Universe. 
These constraints will then be robustly compared to simulations, as they will form a complete $K$-band selected sample.
The statistics will enable to provide tighter constraints to galaxy formation and mass buildup mechanisms at $z>6$, going beyond what has been possible in the pre-JWST era \cite[e.g.,][]{Forrest2020b,DEugenio2021a, Kalita2021a}.
Additionally, early results from JWST Early Release Science imaging have shown evidence for more massive quiescent galaxy candidates at $z>3$ \citep{Carnall2023a} that were not selected by ground-based surveys. 
Spectroscopic confirmation of these sources along with constraints to their formation mechanisms would build up a picture of the early mass buildup of the Universe which can be tested with updated galaxy evolution models utilized by the new generation of simulations \citep[e.g.][]{Kimmig2023a}.

Prominent broad \OIII\ and \Halpha\ emission lines are visible in our galaxies which could be indicative of AGN activity. 
The \OIII/\Hbeta\ vs stellar mass diagnostic from \citep{Juneau2014a} suggests that these emission lines could be powered by an AGN. 
Given the low resolution of the NIRSpec/Prism mode observations, we are unable to separate the broad and narrow components of these emission lines. Thus, we cannot rule out secondary contribution from star-formation to these lines.
Recent results have  shown evidence for AGN to play a prominent role in the evolution of massive quiescent galaxies at $z\sim3-5$. 
For example, NIRSpec medium resolution spectroscopy of a $z\sim4.6$ quiescent galaxy candidate \citep{Carnall2023b} show clear evidence for broad outflows likely driven by an AGN. 
Similarly, the most massive galaxies spectroscopically confirmed at higher redshifts also allude to presence of accreting black holes \citep{Maiolino2023a}. 
In our sample, most of the younger quiescent galaxies do show strong emission line ratios consistent with powered through AGN activity. 
While the post-starburst nature of low mass galaxies at $z>5$ \citep{Looser2023a,Strait2023a} could be driven by the prominence of strong star-bursts, our massive galaxies would likely require more intense feedback mechanisms to shut down star-formation. 
Future work is needed to disentangle the contribution of AGN feedback from the contribution of intense star formation within the context of the new generation of large cosmological simulations to test efficient quenching mechanisms of early massive galaxies \citep[e.g.][]{Kimmig2023a,Schaye2023a}.

Our understanding of the formation of these early quiescent systems will improve rapidly due to the new capabilities of JWST. The new data presented here are only 33 minute integrations of \emph{HST} selected samples at relatively low spectral resolution with a median S/N of $\sim10-70$ and are only a first look. 
Future higher spectral and spatial resolution JWST/NIRSpec observations of these galaxies have the capability to strengthen our understanding of the stellar populations, initial mass functions \citep{Esdaile2021a}, and mass buildup mechanisms and timescales through measurements of their kinematic properties and abundance measurements
of multiple elements \citep{Nanayakkara2021}.
These massive quiescent galaxies provide a fossil record of star-formation in the Universe prior to $z=4$, in systems where nature has helpfully formed an abundance of stars in one place where detailed high S/N spectral analysis with JWST will be possible.

This work is currently limited to presenting a first JWST/NIRSpec view of a \emph{HST} selected population of massive quiescent galaxy candidates. 
The work mirrors the approach taken by S18 to provide a view of the transformative capabilities of JWST. 
A thorough analysis of ZF-UDS-7329 is presented in \cite{Glazebrook2023a} with advanced modelling of parametric and non-parametric star-formation histories. 
Morphological analysis based on $1-5~\mu m$ PRIMER and CEERS imaging of our galaxies rule out any secondary reddened components for our sources (e.g. \cite{Kalita2022a,Chandar2023a}).
A detailed JWST/NIRcam morphological analysis of the S18 sources will be presented by Kawinwanichakij et al. (in prep). 
Nanayakkara et al. (in prep) will present the full S18 sample observed by GO-2565 program upon completion of data acquisition (imaging and spectra) and analysis with complex star-formation modelling including detailed contributions from AGN using {\tt BEAGLE} \citep{Chevallard2016} and {\tt Prospector} \citep{Johnson2021a}.

\section{Figures and Tables}\label{fig:figures}

\begin{sidewaystable}
\footnotesize
\sidewaystablefn%
\begin{minipage}{174pt}
\caption{Galaxy spectroscopic redshifts, NIRCam $F200W$ band magnitudes, and {\tt FAST++} best fit parameters for galaxies presented in this analysis. {\tt FAST++} parameters are as defined as in S18. $1\sigma$ upper and lower bounds for {\tt FAST++} measured values derived using 1000 MCMC iterations are included with the best fit parameters.}\label{tab:fastparams}%
\begin{tabular}{@{}lllllllllll@{}}
\toprule
Galaxy ID & $z_{spec}$&  $F200W^a$ &  $M_*$   & Z           &  Av    & $\mathrm{SFR_{10}^b}$              & $t_{quench}^c$     & $t_{form}^d$       & $t_{\mathrm{SF}}^e$  &  $\langle SFR \rangle_{main}^f$ \\
          &            & AB mag & $\rm{log_{10}}(M_\odot)$ &   & mag     & $\rm{log_{10}}(M_\odot/yr)$  &  $\rm{log_{10}}(yr)$  & $\rm{log_{10}}(yr)$   & $\rm{log_{10}}(yr)$      & $\rm{log_{10}}(M_\odot/yr)$ \\
\midrule
3D-EGS-18996 & $3.250^{3.252}_{3.249}$ & $21.860^{21.860}_{21.860}$ & $10.880^{10.880}_{10.840}$ & $0.050^{0.044}_{0.004}$ & $0.000^{0.000}_{0.000}$ & $-1.290^{-1.290}_{-4.260}$ & $8.000^{8.500}_{8.000}$ & $8.650^{8.730}_{8.640}$ & $8.740^{8.740}_{8.570}$ & $2.160^{2.300}_{2.160}$ \\
3D-EGS-31322 & $3.434^{3.435}_{3.433}$ & $22.480^{22.490}_{22.480}$ & $10.750^{10.750}_{10.740}$ & $0.050^{0.045}_{0.025}$ & $0.300^{0.290}_{0.220}$ & $-4.100^{-4.060}_{-6.070}$ & $8.380^{8.380}_{8.380}$ & $8.450^{8.450}_{8.440}$ & $7.490^{7.500}_{7.460}$ & $3.270^{3.290}_{3.260}$ \\
3D-EGS-34322 & $3.227^{3.231}_{3.223}$ & $24.710^{24.720}_{24.690}$ & $10.160^{10.190}_{10.040}$ & $0.020^{0.019}_{0.006}$ & $1.800^{2.000}_{1.730}$ & $1.090^{1.190}_{-0.280}$ & $7.270^{7.490}_{6.970}$ & $8.100^{8.140}_{7.700}$ & $7.480^{7.770}_{7.390}$ & $2.660^{2.720}_{2.320}$ \\
ZF-UDS-3651 & $3.813^{3.814}_{3.813}$ & $23.440^{23.450}_{23.430}$ & $10.650^{10.660}_{10.640}$ & $0.008^{0.007}_{0.004}$ & $1.300^{1.310}_{1.190}$ & $-3.350^{-1.210}_{-2.800}$ & $8.000^{8.000}_{7.930}$ & $8.090^{8.190}_{8.040}$ & $7.540^{7.680}_{7.200}$ & $3.090^{3.440}_{2.960}$ \\
ZF-UDS-4347 & $3.703^{3.705}_{3.702}$ & $23.550^{23.560}_{23.540}$ & $10.450^{10.450}_{10.420}$ & $0.020^{0.017}_{0.004}$ & $0.400^{0.380}_{0.170}$ & $-6.110^{-4.110}_{-15.850}$ & $8.230^{8.620}_{8.250}$ & $8.530^{8.660}_{8.460}$ & $8.360^{8.360}_{7.590}$ & $2.110^{2.850}_{2.110}$ \\
ZF-UDS-6496 & $3.976^{3.978}_{3.974}$ & $23.170^{23.170}_{23.160}$ & $10.860^{10.880}_{10.860}$ & $0.050^{0.045}_{0.023}$ & $0.000^{0.070}_{0.000}$ & $-2.490^{-2.400}_{-3.900}$ & $8.420^{8.440}_{8.370}$ & $8.790^{8.790}_{8.740}$ & $7.890^{8.690}_{7.800}$ & $3.010^{3.100}_{2.210}$ \\
ZF-UDS-7329 & $3.207^{3.209}_{3.206}$ & $22.830^{22.830}_{22.820}$ & $11.100^{11.120}_{11.060}$ & $0.020^{0.018}_{0.010}$ & $0.300^{0.270}_{0.140}$ & $-2.440^{-0.950}_{-19.930}$ & $9.010^{9.060}_{8.950}$ & $9.160^{9.220}_{9.140}$ & $8.380^{8.710}_{8.310}$ & $2.770^{2.810}_{2.440}$ \\
ZF-UDS-7542 & $3.208^{3.208}_{3.206}$ & $23.410^{23.410}_{23.400}$ & $10.690^{10.700}_{10.660}$ & $0.050^{0.044}_{0.021}$ & $1.100^{1.360}_{0.920}$ & $-1.790^{0.710}_{-2.380}$ & $7.840^{8.150}_{7.830}$ & $8.450^{8.610}_{8.250}$ & $8.730^{8.910}_{7.410}$ & $1.970^{3.240}_{1.810}$ \\
ZF-UDS-8197 & $3.550^{3.550}_{3.549}$ & $23.830^{23.840}_{23.820}$ & $10.400^{10.410}_{10.390}$ & $0.008^{0.007}_{0.004}$ & $1.100^{1.090}_{0.970}$ & $-3.920^{-3.870}_{-6.950}$ & $8.120^{8.230}_{8.110}$ & $8.280^{8.350}_{8.260}$ & $7.860^{7.930}_{7.410}$ & $2.530^{3.010}_{2.470}$ \\
3D-UDS-35168 & $3.529^{3.536}_{3.525}$ & $24.230^{24.240}_{24.230}$ & $10.180^{10.180}_{10.160}$ & $0.008^{0.007}_{0.004}$ & $0.000^{0.000}_{0.000}$ & $-0.270^{-0.190}_{-0.520}$ & $8.400^{8.410}_{8.320}$ & $9.030^{9.050}_{9.030}$ & $8.150^{8.360}_{8.120}$ & $2.070^{2.100}_{1.820}$ \\
3D-UDS-39102 & $3.587^{3.587}_{3.586}$ & $23.270^{23.420}_{23.120}$ & $10.770^{10.850}_{10.740}$ & $0.020^{0.040}_{0.009}$ & $1.800^{1.870}_{1.610}$ & $-1.610^{0.470}_{-1.310}$ & $7.710^{7.750}_{7.490}$ & $7.900^{7.970}_{7.820}$ & $7.360^{7.750}_{7.300}$ & $3.380^{3.460}_{2.980}$ \\
3D-UDS-41232 & $3.121^{3.122}_{3.119}$ & $21.730^{21.750}_{21.720}$ & $11.170^{11.170}_{11.160}$ & $0.050^{0.045}_{0.021}$ & $0.400^{0.390}_{0.310}$ & $-1.180^{0.100}_{-2.150}$ & $8.410^{8.520}_{8.410}$ & $8.790^{8.790}_{8.760}$ & $7.910^{8.160}_{7.850}$ & $3.290^{3.350}_{3.080}$ \\
\end{tabular}
\footnotetext[a]{For galaxies without NIRCam photometry, ground based $K$ band magnitude from the 3DHST survey is presented.}
\footnotetext[b]{SFR of the galaxy computed over a lookback time of 10 Myrs.}
\footnotetext[c]{The lookback time for when the galaxy is considered quenched. }
\footnotetext[d]{The lookback time for when the galaxy formed 50\% of its total stellar mass. }
\footnotetext[e]{The length of time where 68\% of the total integrated SFR of the galaxy took place. This window surrounds the peak SFR of the galaxy. }
\footnotetext[f]{The average SFR in the time window defined by $\mathrm{t_{SF}}$. }
\end{minipage}
\end{sidewaystable}


\clearpage

\section{Methods}

\subsection{Spectrophotometric Calibration}
\label{sec:spec_cal}

The data were reduced using the publicly available pipeline {\tt jwst v1.12.5} with  stages 1, 2, and 3 executed under theJ WST Calibration Reference Data System (CRDS) context {\tt jwst\_1149.pmap} provided by STScI \citep{Bushouse2022a}. The end products of this pipeline are 2D rectified, background-subtracted, and wavelength- and flux-calibrated spectral images, as well as a box car extraction from the 2D product.

A significant limitation of the existing JWST/NIRSpec pipeline arises from the computation of the light {\tt pathloss} function, which accounts for light loss due to the Multi-Shutter Assembly's (MSA) finite slit size (geometrical losses including
those due to the PSF width) and the light dispersion stemming from the instrument's finite pupil size (diffraction loss) \citep{Ferruit2016a}. The 
JWST flight  calibration uses observations of slit-centred
standard stars whose reference spectrum is known. Thus
centred point sources receive an 
empirical absolute spectrophotometric
calibration accounting for these geometric and diffraction
losses. 

For more complex sources the pipeline uses a theoretical
optical model \citep{Ferruit2016a} to calculate the relative pathloss
compared to a centred point source. However the
pipeline only has models for  non-centred point sources
and for sources that uniformly fill a slit.
Considering that our galaxy sample primarily consists of compact galaxies (Kawinwanichakij et al., in prep), neither scenario accurately represents their morphology. Estimating a correction based solely on morphological parameters is complicated and necessitates a model that accounts for the spatial light distribution of the object as a function of wavelength, including PSF effects.

Our first goal is to empirically verify that the
absolute spectrophotometry {\it though the slit}, including PSF effects, 
is accurate. Our logic is that JWST NIRCam images should have a very similar PSF to that seen by NIRSpec as to first order the PSF represents the telescope 
optics diffraction with the different backend instrument optics creating only secondary modifications. Therefore we can add pseudo-slit apertures to NIRCAM
images and compare fluxes in these apertures to the NIRSpec spectra. 
We use the NIRcam Primer images (data obtained from https://dawn-cph.github.io/dja/ v7 data release) in the \emph{F115W, F150W, F200W, F277W, F356W, F410M, F444W} bands and superimpose artificial slits in a 5-slit pattern over 3 dither position on all these images (Kawinwanichakij et al., in prep)  to calculate the total flux falling within the slit aperture.

To obtain our NIRSpec spectrophotometry we take advantage of the fact that by
definition the {\it uniform} pathloss correction file is simply
the inverse of the pathloss correction for a centered point source
(\citep{Ferruit2016a}, p.10). This arises because a uniform source has zero
absolute pathloss. Thus by applying the uniform pathloss one simply
gets the flux through the slit, regardless of image morphology.
To achieve this we
reduce all our spectroscopic data data forcing the NIRSpec pipeline to assume them as point sources for calibration. We further remove the {\tt pathloss} step in the stage 2 pipeline that would correct for de-centering. Then we 
optimally extract 1D spectra and correct these using the uniform pathloss
reference file.  
At each dither position we approximated our sources to be illuminated following a 3-slitlet pattern, given currently there are no JWST reference files for a 5-slitlet slit. We predict the result of this process to agree closely with the PRIMER photometry.

\begin{sidewaysfigure}[h]%
\includegraphics[scale=.3]{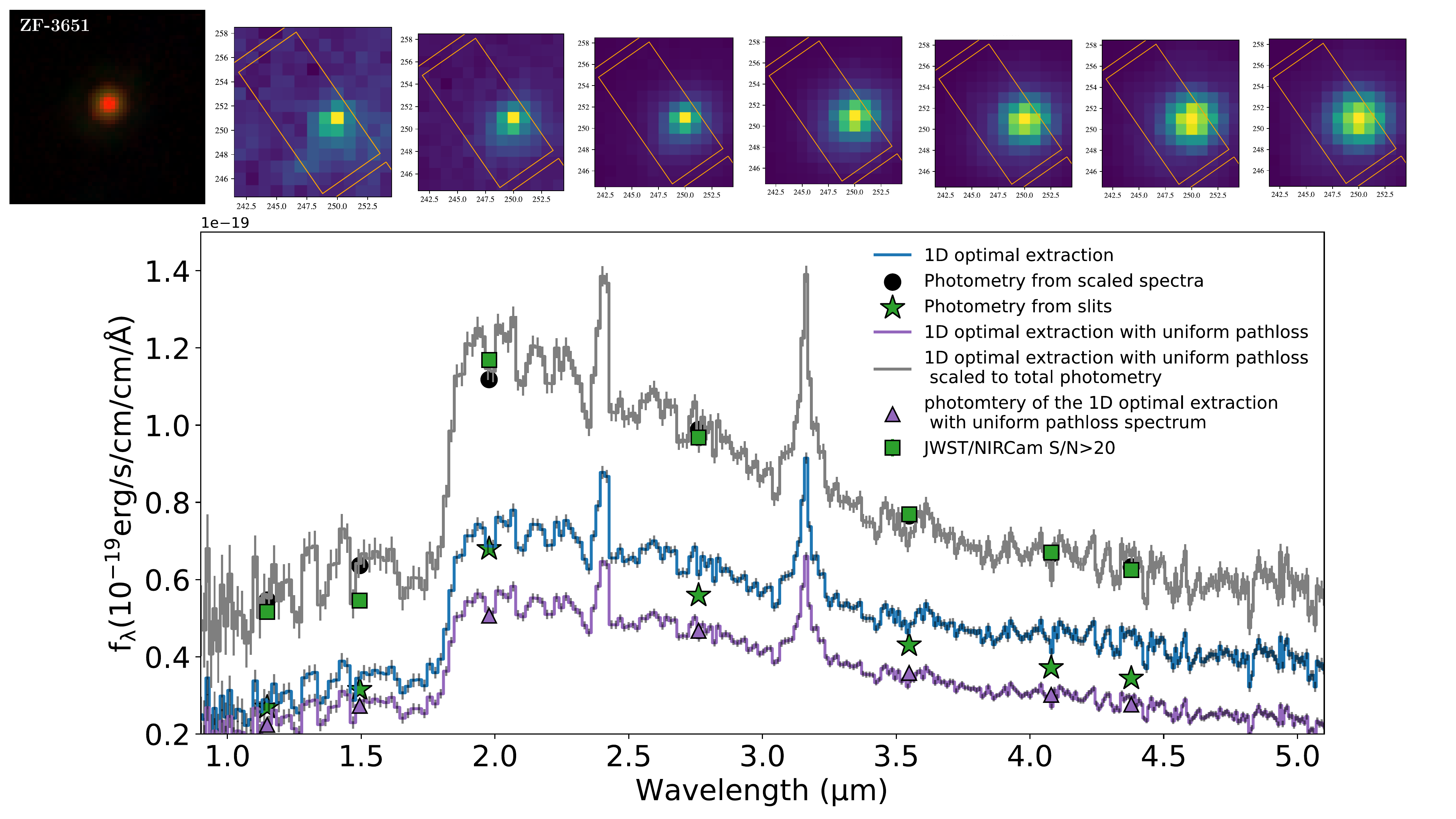}
\caption{An example demonstrating our total flux scaling process for ZF-UDS-3651. In the upper panel we show the NIRCam images of the galaxy. From left to right: [$F115W$, $F200W$, $F444W$] colour composite image covering  rest frame optical bands. The next seven panels are the NIRCam [$F115W$, $F150W$, $F200W$, $F277W$, $F356W$, $F410m$, $F444W$] images. In each panel, the slit at dither position 1 is overlaid on the image. Flux through the slit is computed using the segmentation image of the source at all three dither positions. 
In the lower panel we show the 1D optimally extracted spectra of  ZF-UDS-3651. In blue we show the 1D optimally extracted spectrum from the final 2D result from the JWST calibration pipeline. The spectrum after a uniform pathloss correction is applied is shown in purple. Photometry from the spectrum are shown by the purple triangles and the photometry computed through the slit images of the respective NIRCam bands are shown by the green stars. In grey we show the spectrum scaled to total NIRCam photometry (green squares) and its corresponding photometry in the filter by black circles.  
\label{fig:flux_cal_fig}}
\end{sidewaysfigure}

In Figure \ref{fig:flux_cal_fig} we show this comparison, both before and after uniform pathloss correction is applied for ZF-UDS-3651. This is an ideal test case as this is an extended galaxy that sits near the edge of its slit.
For all our sources, the  RMS between slit vs uniform pathloss photometry is $\sim 5.8 \times 10^{-21} \rm{erg/s/cm^2/}$)].
Note this does not mean it represents the true flux on sky within the aperture because both NIRCam and NIRSpec data now include slit losses from
PSF wings that will increase with wavelength. Any non-uniform source, even one with a uniform colour, will see a small chromatic shift due to the wavelength dependent PSF when viewed through a slit aperture. 
However the fact that NIRCam and NIRSpec agree means we have an understanding of the spectrophotometry of complex  sources and validates our assumption that two instruments have very similar PSFs.

We next determine the correction to total spectrophotometry using the PRIMER photometry. 
First we investigate the degree of any spectral shift introduced by this by using the  $F150W-F444W$ colour as a proxy (as it straddles the 
Balmer breaks in our objects). We find the median shift in colour between slit and total to be at the $\sim 0.05 \pm 0.2$ mag level for all our sources, which we attribute to the relatively homogeneous colours of the quiescent galaxies.

We then used the photometry computed using the spectra to derive an 
empirical polynomial calibration function to match the total photometric magnitude of the objects. 
For this purpose we use all PRIMER photometry with a S/N$>20$. We fit a 2nd order polynomial to the line flux difference and apply that correction to the observed uniform pathloss corrected spectra. Similar to before, we further validate that this process does not introduce an extra color bias to the spectra. We find the offset to be at the $\sim 0.05 \pm 0.1 $ mag level. 
This scaled spectra is then used in spectral fitting using {\tt FAST++}.

\subsection{Spectrophotometric Fitting with {\tt FAST++}}
\label{sec:sed_fitting}

The galaxies studied in this work were selected from S18 for spectroscopic confirmation, and our {\tt FAST++} analysis closely mirrors the SFH analysis of S18. 
However, we allow the stellar metallicity to vary between 20-250\% $Z_{\odot}$ and make minor adjustments to step sizes of the parameters that define the SFH as outlined in Table \ref{table:fastpp_sfh_param}.

S18 tailored the analytical form of the SFH  to specifically suite quiescent galaxies at $z\sim3-4$. 
The elaborate functional form developed by S18 allows the star-forming phase of the galaxies to be broken into two separate epochs. 
The primary phase of the SFH comprise of an exponentially increasing and decreasing window and is expressed as follows: 

\[
SFR_{\text{Base}}(t) \propto \begin{cases} 
e^{(t_{\text{burst}}-t)/\tau_{\text{rise}}} & \text{for } t > t_{\text{burst}}, \\
e^{(t-t_{\text{burst}})/\tau_{\text{decl}}} & \text{for } t \leq t_{\text{burst}},
\end{cases}
\]

where $SFR_{\text{Base}}(t)$ is the base SFR at lookback time t, and $t_{burst}$ defines the boundary between e-folding times of the exponentially increasing ($\tau_{rise}$) and decreasing ($\tau_{decl}$) parts of the SFH. 

Additionally, a scaling factor ($R_{SFR}$) is introduced during the last 10-300 Myr ($t_{free}$) of the galaxy SFH to momentarily increase the SFH near the observation time as follows: 
\[
SFR(t) = SFR_{\text{Base}}(t) \times \begin{cases} 
1 & \text{for } t > t_{\text{free}}, \\
R_{SFR} & \text{for } t \leq t_{\text{free}}.
\end{cases}
\]
where $SFR(t)$ is the final SFR of the source at lookback time $t$. This adjustment is crucial for accounting for recent SFHs in otherwise quenched galaxies, especially when considering star formation rate (SFR) measurements derived from FIR observations. 
We refer the readers to \cite{Schreiber2018,Schreiber2018b} for further information on the SFH.

\begin{figure}[h]%
\includegraphics[scale=0.75]{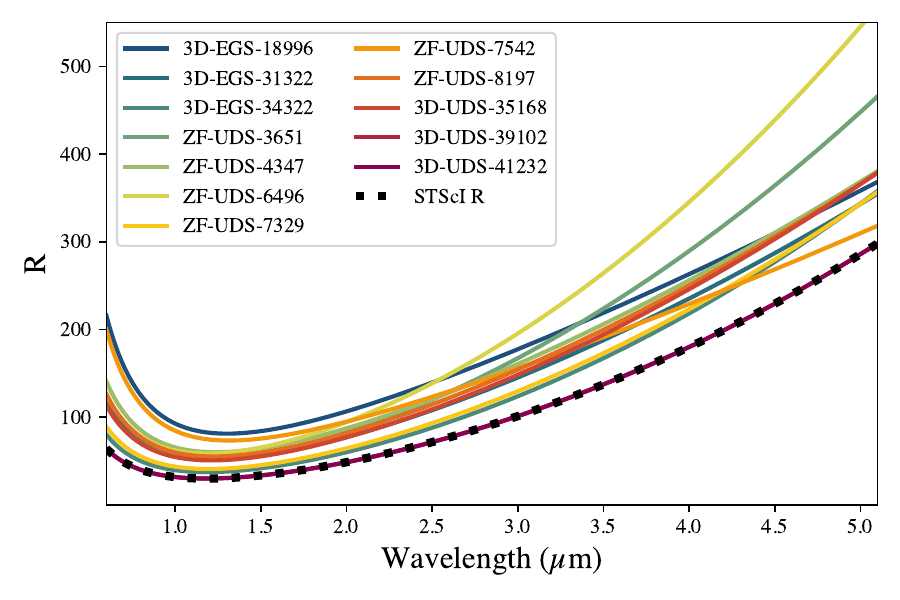}
\caption{The observed NIRSpec/Prism mode spectral resolution for our sample. These are empirically computed by modelling the image size on the NIRSpec MSA slit using PRIMER NIRCam imaging. 
The spectral resolution of the NIRSpec/Prism mode for a uniformly illuminated slit is shown by the black dashed line for comparison. Given our objects do not fill the full slit width, the observed resolution is higher than the values presented by STScI. 
\label{fig:lsfs}}
\end{figure}

NIRSpec/Prism mode has a non-linear spectral resolution and dispersion. Thus, for accurate spectral fitting we need to convolve the model spectra with an accurate dispersion based on the source profile on the slit. 
We use the $1-5~\mu m$ NIRCam images from PRIMER to compute an empirical function on how the FWHM of the source image vary on the slit as a function of wavelength. This is well approximated by a linear function for our sources. We then multiply this function by the NIRSpec/Prism dispersion provided by STScI (https://jwst-docs.stsci.edu/jwst-near-infrared-spectrograph/nirspec-instrumentation/nirspec-dispersers-and-filters\#NIRSpecDispersersandFilters-DispersioncurvesfortheNIRSpecdispersers; also \cite{Jakobsen2022a}) to obtain the LSF of the galaxies. This is then used by {\tt FAST++} to convolve the model spectra to match with the observations. The spectral resolution computed for our sources are shown by Figure \ref{fig:lsfs} and more details are presented in \cite{Glazebrook2023a}. For the two galaxies with no NIRCam imaging, we use the LSF of a uniformly illuminated slit, as provided by STScI.

\begin{table}[ht]
\centering
\caption{The SFH parameters used in {\tt FAST++}.}
\label{table:fastpp_sfh_param}
\begin{tabular}{l rc c c }
\hline
Free Parameter & Lower Bound & Upper Bound & Step Size \\
\hline
$t_{burst}$ (Gyr)  & 0.01 & tH(z) & 0.025 dex   \\
$\tau_{rise}$ (Gyr) & 0.01 & 3 & 0.25 dex   \\
$\tau_{decl}$ (Gyr) & 0.01 & 3 & 0.05 dex   \\
$R_{SFR}$          & $10^{-2}$ & $10^5$ & 0.1 dex \\
$t_{free}$ (Myr)   & 10 & 300 & 0.25 dex   \\
\hline
\end{tabular}
\end{table}

\backmatter

\bmhead{Acknowledgments}
This work is based on observations made with the NASA/ESA/CSA James Webb Space Telescope. The data were obtained from the Mikulski Archive for Space Telescopes at the Space Telescope Science Institute, which is operated by the Association of Universities for Research in Astronomy, Inc., under NASA contract NAS 5-03127 for JWST. These observations are associated with program 2565. We thank all the hard work of the JWST team which made this great observatory possible.
We thank Michael Maseda and Allison Strom for helpful discussions during the data reduction process.
T.N., K. G., and C.J. acknowledge support from Australian Research Council Laureate Fellowship FL180100060. 
This work has benefited from funding from the Australian Research Council Centre of
Excellence for All Sky Astrophysics in 3 Dimensions (ASTRO 3D), through project number CE170100013.
The Cosmic Dawn Center is funded by the Danish National Research Foundation (DNRF) under grant DNRF140. 
P.O. is supported by the Swiss National Science Foundation through project grant 200020\_207349. This work received funding from the Swiss State Secretariat for Education, Research and Innovation (SERI).

\bmhead{Data Availability}
Data used in the analysis is available in the NASA Mikulski Archive for Space Telescopes (MAST) under program ID 2565  DOI: 10.17909/trb7-hs97. Alternatively the data is also available at The DAWN JWST Archive https://dawn-cph.github.io/dja/index.html under Program ID 2565 Observations 100 and 200. We thank the curators of both services for hosting the data. All high-end data products can also be provided by the lead author on reasonable request. 

\bmhead{Code Availability}
All softwares used in this analysis are publicly available.


\bibliography{bibliography_qu_prism}

\end{document}